\documentclass[epj,twocolumn]{webofc}
\usepackage[varg]{txfonts}   % Web of Conferences font
\usepackage{epstopdf}

\woctitle{LHCP 2013}
\begin{document}
%\linenumbers
%
\title{BSM Higgs results from ATLAS and CMS}

\author{Martin Flechl\inst{1}\fnsep\thanks{\email{martin.flechl@cern.ch}} on behalf of the ATLAS and CMS collaborations}

\institute{Albert-Ludwigs-Universit\"at Freiburg, Physikalisches Institut, Hermann-Herder-Str. 3, 79104 Freiburg, Germany}

\abstract{%
Searches for Higgs bosons in different extensions of the Standard Model (SM) are presented. These include the 
Minimal Supersymmetric extension of the SM (MSSM), the next-to-MSSM (NMSSM), 
models with additional scalar singlets, doublets, or triplets, and generic searches for models with 
couplings modified with respect to the SM or for non-SM Higgs boson decay channels. Results are based on data 
collected by the ATLAS and CMS experiments in 2011 and 2012 at the LHC. No excess is found in any of the searches 
and thus the resulting exclusion limits are given.
}
\maketitle
\section{Introduction}\label{intro}
A large number of searches for beyond-the-SM (BSM) Higgs bosons have been performed by the ATLAS~\cite{atlas} and CMS~\cite{cms} experiments 
at the LHC~\cite{lhc}, using collision data at a center-of-mass energy of 7 TeV (2011) and 8 TeV (2012). 
The discovery of a particle with a mass of 125~GeV compatible with a Higgs boson by the ATLAS and CMS experiments~\cite{disc_atlas,disc_cms} 
leads to new questions -- in particular whether it is \textit{the} SM Higgs boson or one of typically many Higgs bosons 
predicted by extensions of the Standard Model. Experimentally, this implies measuring the properties of the new boson 
(not covered here) and searching for additional Higgs bosons. In the following an overview of the most recent results of searches for BSM Higgs bosons 
using an integrated luminosity of at least 5~fb$^{-1}$ is presented.

\vspace*{1.0cm}
\section{MSSM Higgs boson searches}\label{sec-2}
The MSSM employs a type-II 2-Higgs doublet model (2HDM), which predicts the existence of three neutral Higgs bosons ($h$, $H$, $A$) 
and a charged pair ($H^\pm$). At tree-level, the Higgs sector of the MSSM can be described by only two additional parameters, 
e.g. by one of the masses of the heavy Higgs bosons and $\tan \beta$, the ratio of the vacuum expectation values of the two 
Higgs doublets. In the MSSM, the recently discovered boson with $m \approx$ 125~GeV can be identified with the $h$ boson for 
most of the $m_{A} - \tan\beta$ region not excluded by direct MSSM Higgs boson 
searches and also with the $H$ particle if the charged Higgs boson mass is below 150~GeV and $\tan \beta$ is small~\cite{carena}. 
Since the couplings of the MSSM Higgs bosons to $b$ and $t$ quarks and tau leptons are enhanced for large $\tan \beta$ most 
searches for these bosons focus on Higgs boson production in association with $b$ and $t$ quarks and decay modes 
involving $b$, $t$ and $\tau$ particles.

\subsection{$H^+ \to c\bar{s}$}\label{sec-2_1}
\begin{figure}
\centering
\includegraphics[width=8.5cm,clip]{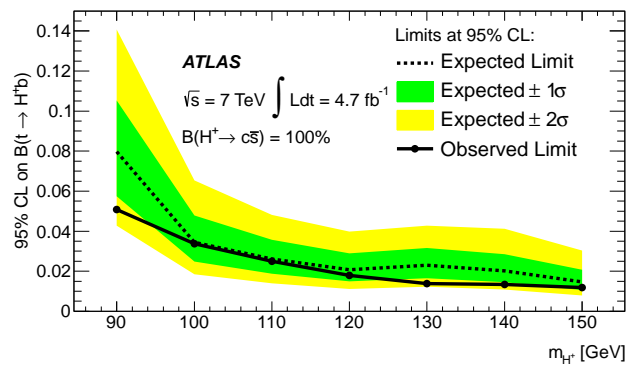}
\caption{ATLAS exclusion limit on B($t \to bH^+$) for charged Higgs boson decays to $c\bar{s}$~\cite{csbar_atlas}, 
assuming B($H^+ \to c\bar{s}$)=$100\%$.}
\label{fig-cs}
\end{figure}

ATLAS searches for charged Higgs bosons produced in semileptonic $t\bar{t}$ events with $t \to bH^\pm$ and $H^\pm \to 
c\bar{s}$ in 4.7~fb$^{-1}$ of 7 TeV data~\cite{csbar_atlas}. 
The analysis requires an isolated lepton, 4 jets (two of them $b$-tagged), missing transverse energy 
($E_T^{\mathrm{miss}}$) greater than 20 (30)~GeV and the transverse mass $m_T$ of the lepton and $E_T^{\mathrm{miss}}$ 
to satisfy $m_T > 30$~GeV ($m_T + E_T^{\mathrm{miss}} > 60$~GeV) in the muon (electron) channel. A kinematic fit is applied to the 
remaining events with the goal to find a second mass peak in the dijet distribution due to $H^+ \to c\bar{s}$ events.
Good agreement between data and SM expectation is observed, leading to a limit on the branching ratio B($t \to bH^+$)=$(5-1)\%$ for 
$m_{H^\pm}=90-150$~GeV, assuming B($H^+ \to c\bar{s}$)=$100\%$, as shown in figure~\ref{fig-cs}.

\subsection{$H^+ \to \tau\nu$}\label{sec-2_2}
Both ATLAS and CMS search for charged Higgs bosons produced in $t\bar{t}$ events that decay to $H^+ \to \tau\nu$. The ATLAS 
analysis~\cite{taunu_atlas} is based on 4.6~fb$^{-1}$ of 7 TeV data and studies the final states $\tau_\mathrm{lep}$+jets, $\tau_\mathrm{had}$+lepton and 
$\tau_\mathrm{had}$+jets where the subscript indicates whether the tau lepton decays to leptons or hadrons+neutrino. 
The $\tau_\mathrm{had}$+jets channel is the most sensitive one for most of the $H^+$ mass range. The selection starts 
with a $\tau+E_T^{\mathrm{miss}}$ trigger and requires the presence of exactly one hadronically decaying tau lepton, 
at least four jets (at least one of them $b$-tagged), no electrons or muons, $E_T^{\mathrm{miss}}>65$~GeV, a high significance of 
the missing transverse energy (quantified by its ratio to the square root of the sum of transverse momenta of reconstructed tracks)
and a topology consistent with the presence of a hadronic top quark decay. The final discriminant used is a transverse mass 
built from the visible tau lepton decay products and the missing momentum vector. All backgrounds are estimated in a data-driven way 
by replacing muons in $\mu$+jets collision data with simulated tau leptons for backgrounds with actual $\tau$ leptons, 
and by deriving data-driven corrections for the misidentification probabilities for backgrounds where jets or electrons are misidentified 
as $\tau$ leptons.   

\begin{figure}
\centering
\includegraphics[width=8.5cm,clip]{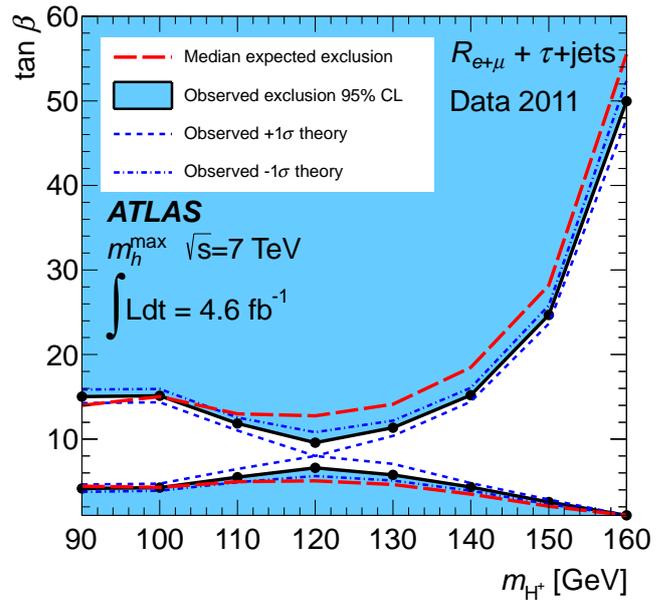}
\caption{ATLAS exclusion limit in the $m_{H^\pm} - \tan\beta$ plane for the $m_h^\mathrm{max}$ scenario of the MSSM 
from the search for charged Higgs boson production in $t\bar{t}$ events and decays to $\tau\nu$. The limits are derived 
using $\tau_\mathrm{had}$+jets events and studying lepton universality~\cite{taunuratio_atlas}.}
\label{fig-hptaunu}       % Give a unique label
\end{figure}

For all final states, the data agree with the SM expectation and an upper limit of B($t \to bH^+)=(5-1)\%$ 
assuming  B($H^+ \to \tau\nu)=100\%$ is set for 
$m_{H^\pm}=90-160$~GeV. In a different approach, lepton universality in $t\bar{t}$ events is tested by observing the ratio of event 
yields with electrons or muons and with and without tau leptons in the final state which would be enhanced by the presence of a charged 
Higgs boson~\cite{taunuratio_atlas}. This brings down the limit after combining with the $\tau_\mathrm{had}$+jets final state to 
about B$(t \to bH^+)=(3-1)\%$. Interpreting this limit in the context of the $m_h^\mathrm{max}$ of the MSSM~\cite{mhmax}, a large portion of the 
parameter space for a light $H^\pm$ can be excluded, see figure~\ref{fig-hptaunu}.

The CMS search~\cite{taunu_cms} uses $2.2-4.9$~fb$^{-1}$ of 7 TeV data in the final states: $\tau_\mathrm{had}$+lepton, $e$+$\mu$, and $\tau_\mathrm{had}$+jets. 
Good agreement between data and expectation is observed and upper limits of B$(t \to bH^+)=(3-2)\%$ for $m_{H^\pm}=80-160$~GeV are set.

\subsection{$\Phi \to bb$}\label{sec-2_3}
\begin{figure}
\centering
\includegraphics[width=8.5cm,clip]{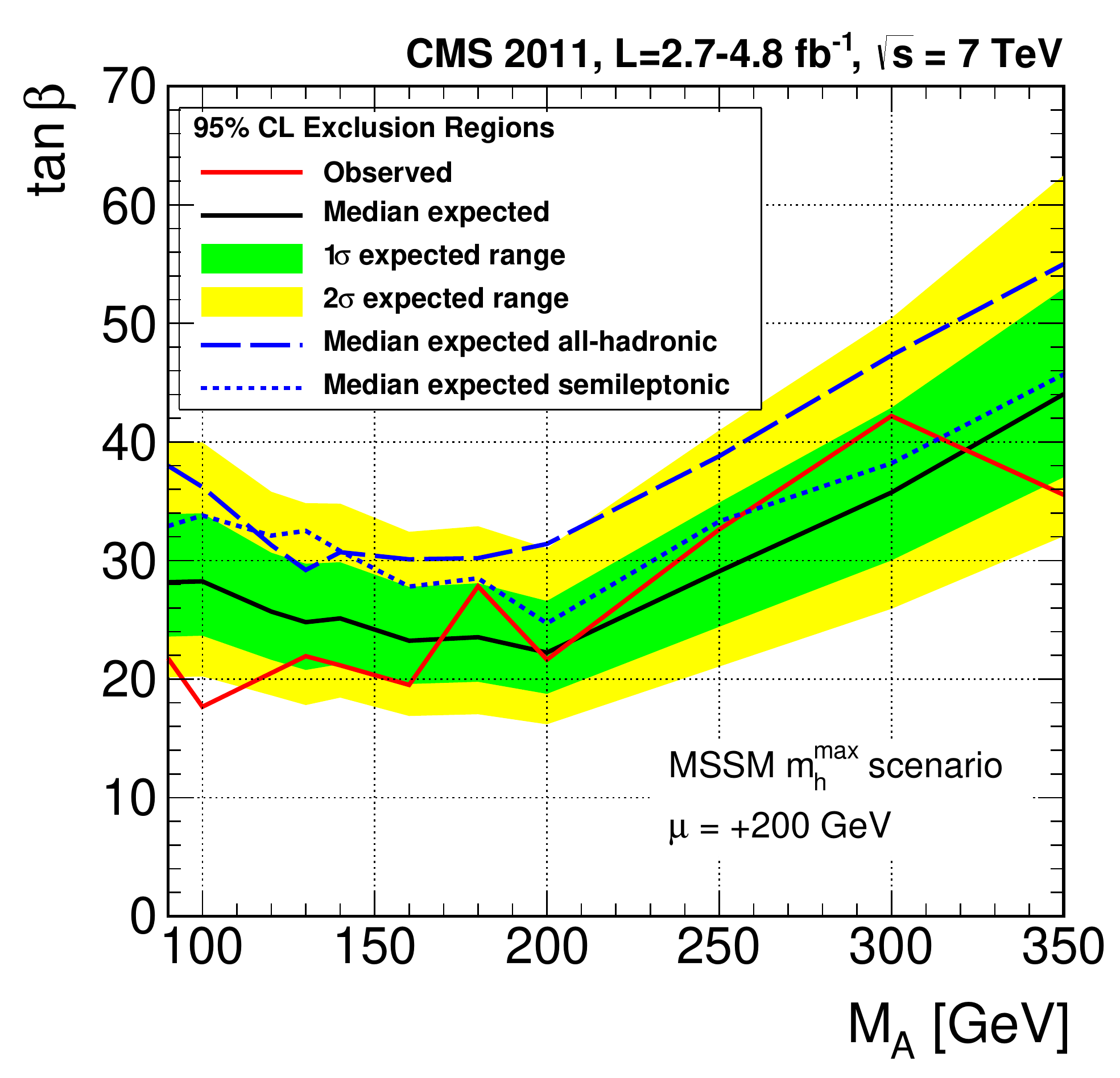}
\caption{CMS exclusion limit for Higgs boson decays via $H \to bb$ in $b$-associated production in the $m_h^\mathrm{max}$ scenario of the MSSM~\cite{bb_cms}.}
\label{fig-bb}
\end{figure}

In the MSSM, the mode $\Phi \to bb$ ($\Phi=h$, $H$, $A$) is dominant for most regions of the parameter space. The CMS search~\cite{bb_cms}
uses up to 4.8~fb$^{-1}$ of 7 TeV data and focuses on $b$-associated production, $pp \to b\Phi \to bbb$, in two separate channels: 
The first requires three tight $b$-tagged jets and is split into two analyses searching for $m_\Phi$ either below or above 180~GeV. 
The second employs looser $b$-tagging requirements but demands that a muon overlaps with the leading 
$b$-tagged jet. The final discriminant is the invariant mass of the two leading $b$-tagged jets. Backgrounds are estimated from 
data, using control samples with 1 or 2 $b$-tagged jets. No excess is observed and combined limits 
exclude values of $\tan \beta > 20-30$ for $m_A=90-350$~GeV in the MSSM ($m_h^\mathrm{max}$ scenario), as shown in figure~\ref{fig-bb}.

\subsection{$\Phi \to \tau\tau$, $\Phi \to \mu\mu$}\label{sec-2_4}
The CMS analysis uses 17~fb$^{-1}$ of 7 TeV and 8 TeV data to search for MSSM Higgs boson decays to a tau lepton pair~\cite{tautau_cms}, and 
5~fb$^{-1}$ of 7 TeV data for decays to a muon pair~\cite{mumu_cms}. For the former, three different combinations of tau lepton decays are considered: 
$\tau_\mathrm{lep} \tau_\mathrm{had}$,  $\tau_\mathrm{e} \tau_\mathrm{\mu}$ and  $\tau_\mathrm{\mu} \tau_\mathrm{\mu}$. The ATLAS 
analysis~\cite{tautau_atlas} in addition studies the $\tau_\mathrm{had} \tau_\mathrm{had}$ final state, but does not use 
$\tau_\mathrm{\mu} \tau_\mathrm{\mu}$. For both ATLAS and CMS, each final state is split into two categories with and without additional $b$-tagged jets 
aiming to exploit the different production mechanisms ($b$-associated and $gg$-fusion). The $b$-tagged categories provide a higher signal-over-background 
ratio but a smaller number of expected signal events. 
The final discriminant in all cases is an estimator, ``Missing Mass Calculator'' for ATLAS and ``SVFit'' for CMS, for the ditau mass 
based on different algorithms for finding the most likely value given the unknowns due to the neutrinos in the decay chain.
The main background in all categories are $Z \to \tau \tau$ events which are estimated using a 
technique replacing muons in $Z \to \mu \mu$-enhanced collision data by simulated tau leptons. Most other relevant backgrounds are normalized in control 
regions.
\begin{figure}
\centering
\includegraphics[width=8.5cm,clip]{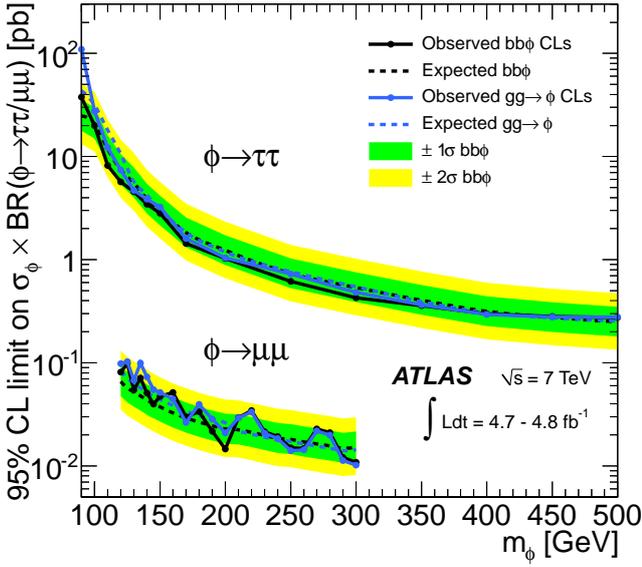}
\caption{ATLAS cross section times branching ratio limits for $gg$-fusion and $b$-associated production of Higgs bosons and subsequent decay 
into tau lepton or muon pairs in the $m_h^\mathrm{max}$ scenario of the MSSM~\cite{tautau_atlas}.}
\label{fig-htautau_atlas}       % Give a unique label
\end{figure}

\begin{figure}
\centering
\includegraphics[width=8.5cm,clip]{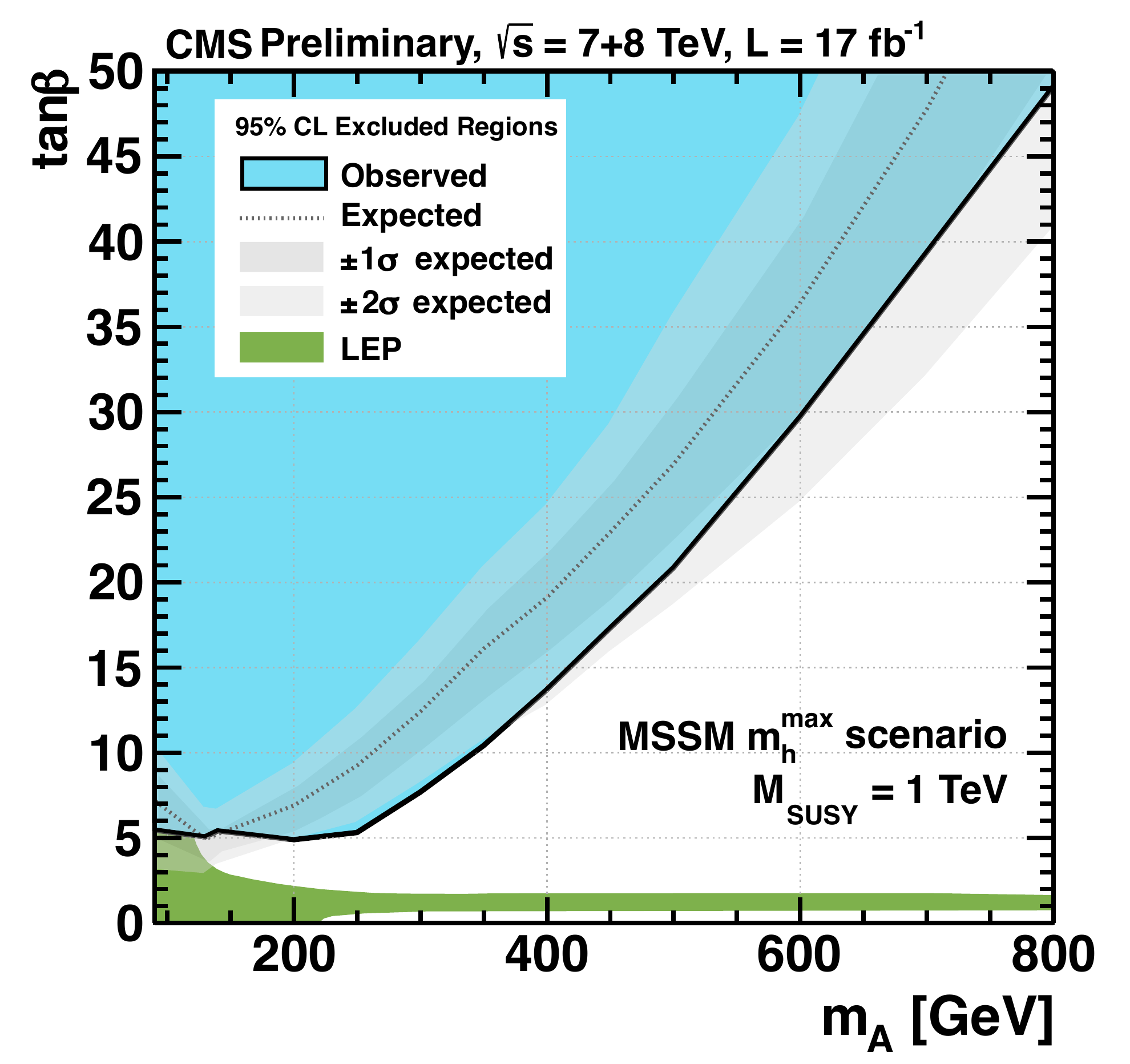}
\caption{CMS exclusion limits for $\Phi \to \tau\tau$ in the $m_h^\mathrm{max}$ scenario of the MSSM~\cite{tautau_cms}. The LEP limit is 
shown as well.}
\label{fig-htautau_cms}
\end{figure}

No significant excess is observed and exclusion limits are produced. ATLAS provides model-independent cross section limits for both gluon-gluon fusion 
and $b$-quark associated production (see figure~\ref{fig-htautau_atlas}); and both experiments interpret their results in the context of 
the $m_h^\mathrm{max}$ scenario of the MSSM (the CMS limit is shown in figure~\ref{fig-htautau_cms}). CMS excludes $m_A<125$~GeV (in the region 
not yet excluded by LEP) as well as values of $\tan \beta>5$ for $m_A<225$~GeV; the exclusion region extends up to $m_A=800$~GeV and $\tan \beta>50$.

\subsection{$H \to WW$}\label{sec-2_5}
\begin{figure}
\centering
\includegraphics[width=8.5cm,clip]{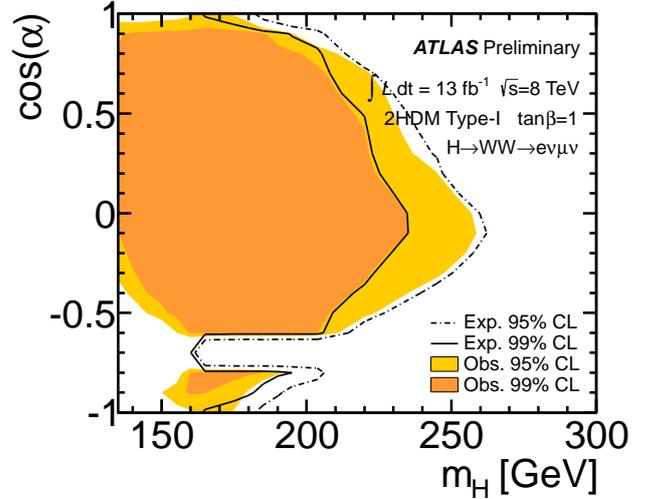}
\caption{ATLAS exclusion limits for $H \to WW$ in a type-I 2HDM for $\tan \beta=1$~\cite{2hdm_atlas}.}
\label{fig-2hdm}
\end{figure}

ATLAS searches for additional CP-even neutral Higgs bosons in a 2HDM, assuming that the observed boson with a mass of 125~GeV is the lightest of the three 
neutral Higgs bosons predicted by a 2HDM. The search focuses on the $H \to WW \to e\nu\mu\nu$ decay mode with 0 or 2 additional jets and uses 
13~fb$^{-1}$ of 8 TeV data~\cite{2hdm_atlas}. The preselection is in common with the SM $H \to WW$ search. To maximize sensitivity, 
an artificial neural net is employed using kinematic information about the event. No evidence for a Higgs boson in the mass range of 
135-300~GeV is found. Exclusion limits are set for type-I and type-II 2HDM for a set of fixed values of $\tan \beta$ ranging from 1 to 50 in the 
$m_H$-$\cos \alpha$ plane 
(an example is shown in figure~\ref{fig-2hdm}). Assuming that other Higgs bosons or additional particles are too heavy to interfere the 
tree-level cross section depends only on these three unknown parameters.

\vspace*{1.0cm}
\section{Generic and exotic Higgs boson searches}\label{sec-3}

\subsection{Heavy Higgs, $H \to VV$}\label{sec-3_1}
Several BSM models are consistent with a Higgs boson with $m=125$~GeV 
and additional heavy Higgs bosons; in the most simple case, only one additional 
real Higgs singlet is added to the Standard Model leading to a second heavy 
Higgs boson which can also be SM-like. Such Higgs bosons are searched for 
in decays to weak bosons, $H \to WW$ and $H\to ZZ$, and the analysis 
strategy closely follows the SM Higgs boson searches.

\begin{figure}
\centering
\includegraphics[width=8.5cm,clip]{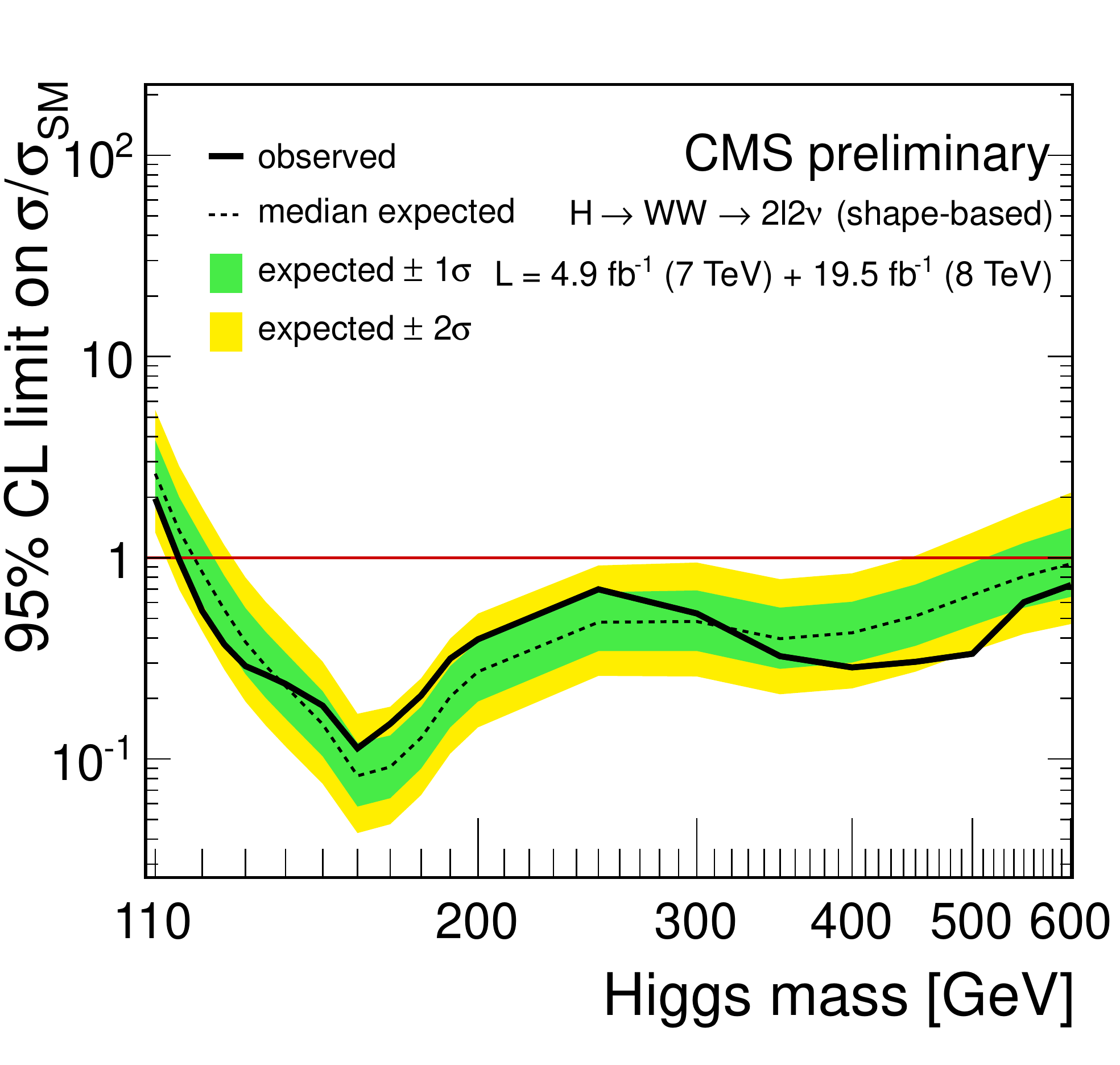}
\caption{CMS exclusion limit for a heavy Higgs boson decaying via $H \to WW \to l \nu l \nu$~\cite{heavyh_wwll_cms}. 
Here, an SM Higgs boson with $m_H$=125~GeV has been added as background process. The limit is given 
in units of the expected SM Higgs cross section as a function of the heavy Higgs boson mass.}
\label{fig-ww}
\end{figure}

CMS uses up to 25~fb$^{-1}$ of 7 TeV and 8 TeV data in searches 
for $H \to WW \to l \nu l \nu$~\cite{heavyh_wwll_cms} and 
$H \to WW \to l \nu qq$~\cite{heavyh_comb_cms}, ATLAS 5~fb$^{-1}$ 
of 7 TeV data in the $l \nu l \nu$ final state~\cite{heavyh_wwll_atlas}. 
No significant excess is observed, and an additional heavy Higgs boson 
with SM couplings is excluded up to a mass of $m_H=600$~GeV; however, 
a cross section times branching ratio less than $10-70\%$ 
(depending on $m_H$) of the 
SM prediction is still allowed, see figure~\ref{fig-ww}. 

\begin{figure}
\centering
\includegraphics[width=8cm,clip]{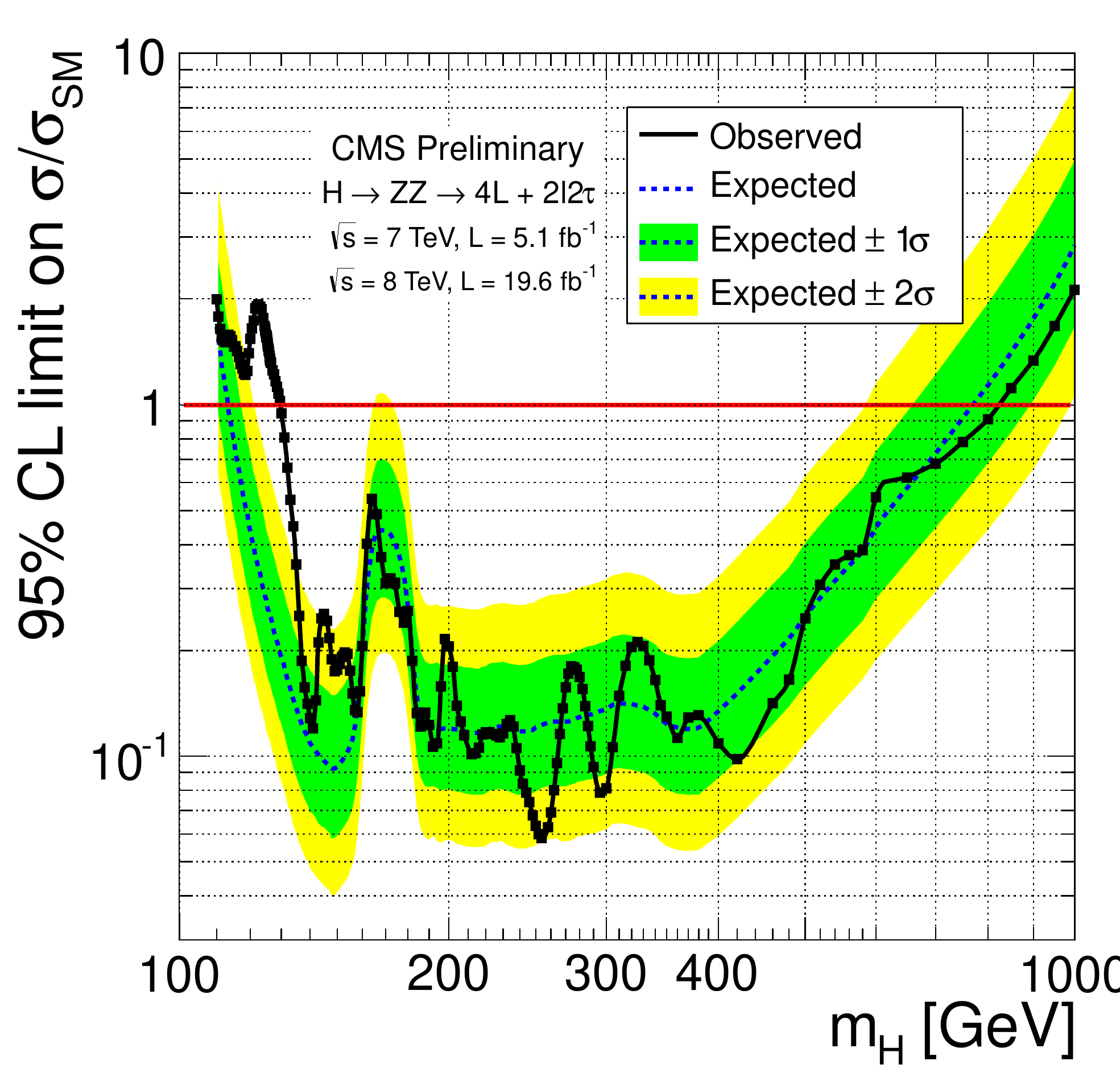}
\caption{CMS exclusion limit for a Higgs boson decaying via $H \to ZZ \to 4l+2l2\tau$~\cite{heavyh_zz4l_lltt_cms}. 
The limit is given in units of the expected SM Higgs cross section as a function of the Higgs boson mass.}
\label{fig-zz}
\end{figure}

Recent heavy Higgs boson searches in the $H \to ZZ$ channel focus on the final states
$2l 2q$~\cite{heavyh_zzllqq_cms}, $2l 2\nu$~\cite{heavyh_comb_cms} and 
$4l+2l2\tau$~\cite{heavyh_zz4l_lltt_cms} for CMS, and $4l$~\cite{heavyh_zz4l_atlas} for 
ATLAS. The data agree well with the SM expectation, and the most stringent limits 
can be set in the $4l+2l2\tau$ analysis and are shown in figure~\ref{fig-zz}. A 
second Higgs boson with SM couplings is excluded for the mass range $m_H=(130-827)$ 
GeV and for $m_H=(180-500)$~GeV, a cross section times branching ratio larger than $20\%$ 
of the SM prediction is excluded.

\subsection{Invisible Higgs, $ZH$}\label{sec-3_2}
\begin{figure}
\centering
\includegraphics[width=8.5cm,clip]{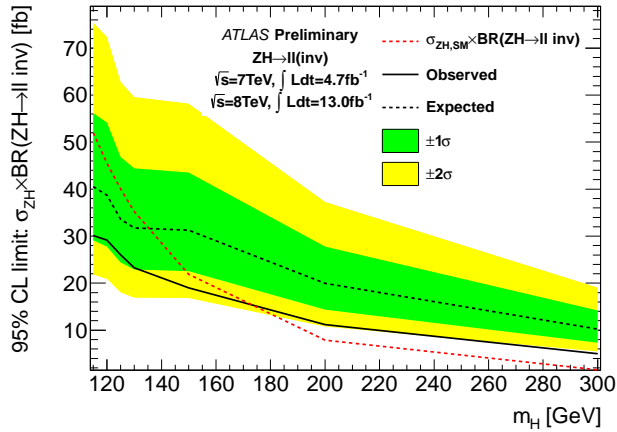}
\caption{ATLAS exclusion limit for invisible Higgs boson decays in $ZH$ production with $Z \to ll$~\cite{zhinv_atlas}. 
}
\label{fig-inv}
\end{figure}

The ATLAS search for invisible Higgs boson decays~\cite{zhinv_atlas} 
considers Higgs bosons produced in association with a $Z$ boson in 18~fb$^{-1}$ of 7 TeV and 
8 TeV data. To tag the event, an electron 
or muon pair consistent with the $Z$ boson mass is required. Events with additional leptons or 
jets are rejected. The missing transverse energy is used as final discriminant. No excess is 
observed, and assuming 
$m_H=125$~GeV and the SM cross section for $ZH$ production, a branching ratio for invisible 
Higgs boson decays larger than $65\%$ is excluded. In addition, upper limits between $30-7$~fb are 
set on the cross section times branching ratios for invisible Higgs boson decays in the mass range $m_H=(115-300)$~GeV, 
see figure~\ref{fig-inv}.

\subsection{$H \to a_0 a_0 \to 4\gamma$}\label{sec-3_3}
\begin{figure}
\centering
\includegraphics[width=8.5cm,clip]{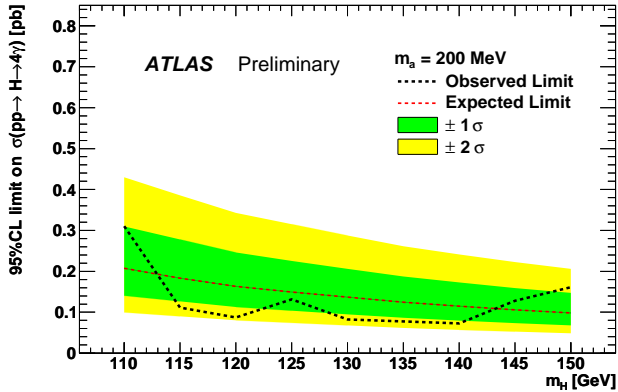}
\caption{ATLAS exclusion limit on the cross section times branching ratio for the process $H \to a_0 a_0 \to 4\gamma$ with $m_{a_0}=200$ MeV. 
Limits for other $m_{a_0}$ masses can be found in reference~\cite{aa_4gam_atlas}.}
\label{fig-aa}
\end{figure}

ATLAS uses 4.9~fb$^{-1}$ of 7 TeV data to search for Higgs boson decays to very 
light CP-odd scalars ($m_{a_0}$ of a few hundred MeV) which in turn decay to photon 
pairs~\cite{aa_4gam_atlas}. This is motivated by composite-Higgs models or the NMSSM 
but the results are interpreted in a model-independent way. Such a process leads to 
two highly collimated $a_0 \to \gamma \gamma$ decays, which can mimick a diphoton 
event. The analysis requires such a diphoton pair using looser requirements for the 
shower shapes for photon identification than e.g. in the SM $H \to \gamma \gamma$ 
search due to two overlapping electromagnetic showers for the signal hypothesis. 
No significant excess is observed and upper limits of a few times 0.1 pb on the 
cross section times branching ratios is set for the mass range $m_H=(110-150)$~GeV 
and $m_A$=100, 200, or 400 MeV. The limit for $m_A$=200 MeV is shown in 
figure~\ref{fig-aa}.
 
\vspace*{0.5cm}
\subsection{Doubly-charged Higgs $\Phi^{\pm\pm}$}\label{sec-3_4}
\begin{figure}
\centering
\includegraphics[clip=true,trim=20cm 0cm 1cm 0cm,width=8cm]{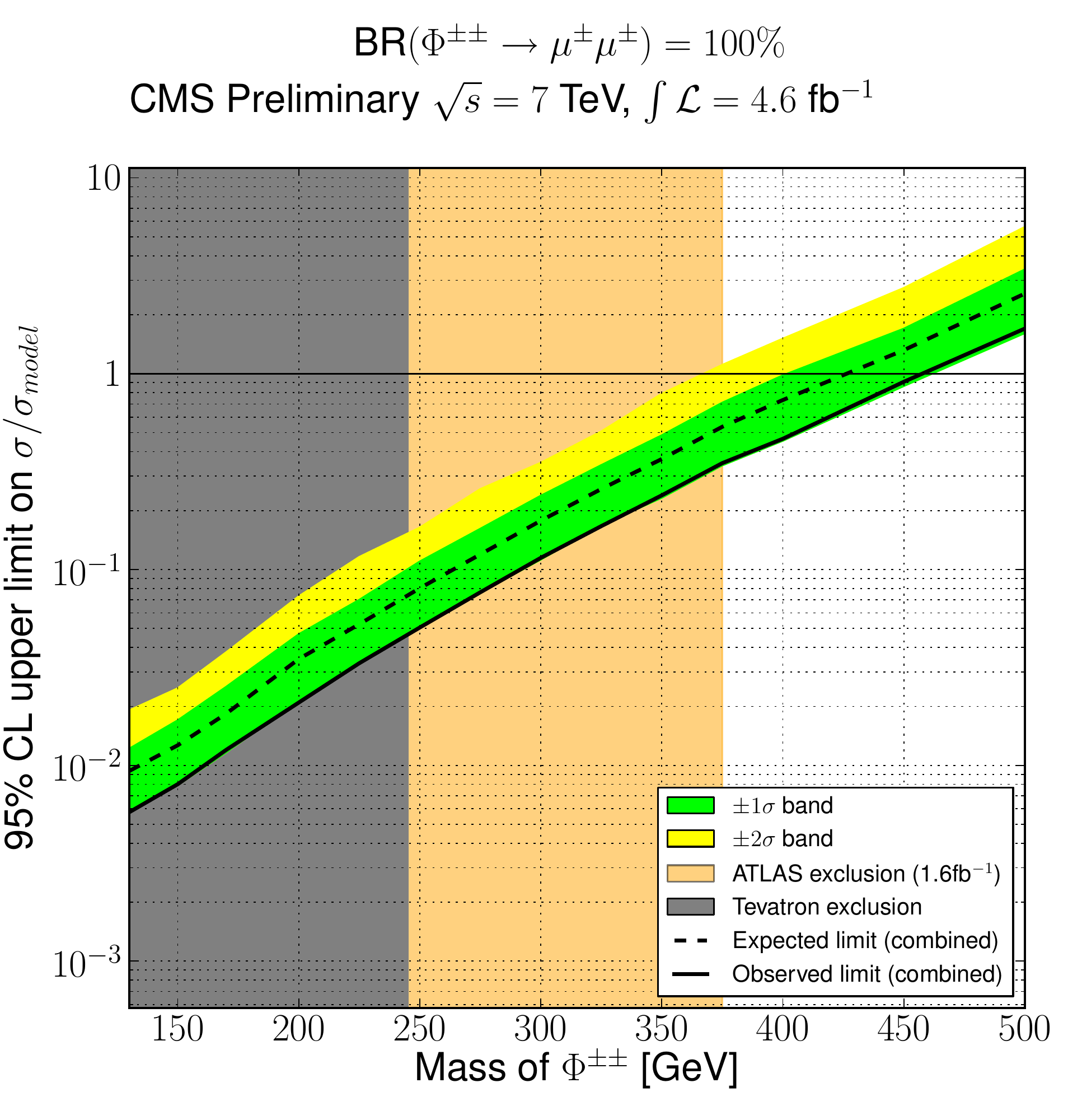}
\caption{CMS exclusion limit for doubly charged Higgs bosons~\cite{doubly_cms}. The combined limit assuming BR($\Phi^{\pm\pm}\to\mu^\pm \mu^\pm$)=100\% is shown. The limit is given in units of the expected $\Phi^{\pm\pm}$ cross section in the investigated type-II seesaw mechanism model.}
\label{fig-doubly}
\end{figure}

Both ATLAS and CMS use about 5~fb$^{-1}$ of 7 TeV data to search for doubly-charged Higgs bosons in decays to 
same-sign lepton pairs $e^\pm e^\pm$, $e^\pm \mu^\pm$, $\mu^\pm \mu^\pm$. CMS, in addition, also looks for all 
possible combinations with tau leptons.
The CMS analysis~\cite{doubly_cms} investigates both pair production of and associated production with a singly-charged 
Higgs boson. The data agree with the SM expectation and upper limits on the cross section times 
branching ratio for $m_{\Phi^{\pm\pm}}=130-500$~GeV are reported for different assumptions on the branching ratio 
of the doubly-charged Higgs boson. This result is translated in lower bounds on $m_{\Phi^{\pm\pm}}$, see figure~\ref{fig-doubly} 
for an example. For a branching ratio of 1 for the different lepton pair decays, doubly-charged Higgs boson 
masses between 204 and 459~GeV can be excluded.

The ATLAS analysis~\cite{doubly_atlas} searches for a narrow peak in the mass spectrum of same-sign lepton pairs. 
No excess is observed and the resulting limits on the cross section times branching ratio lead to 
lower bounds of about 400~GeV on the $\Phi$ mass, assuming pair producion and a branching ratio of 1 to the investigated final states.

\subsection{Higgs in fermiophobic and SM4 models}\label{sec-3_5}
\begin{figure}
\centering
\includegraphics[width=8.5cm,clip]{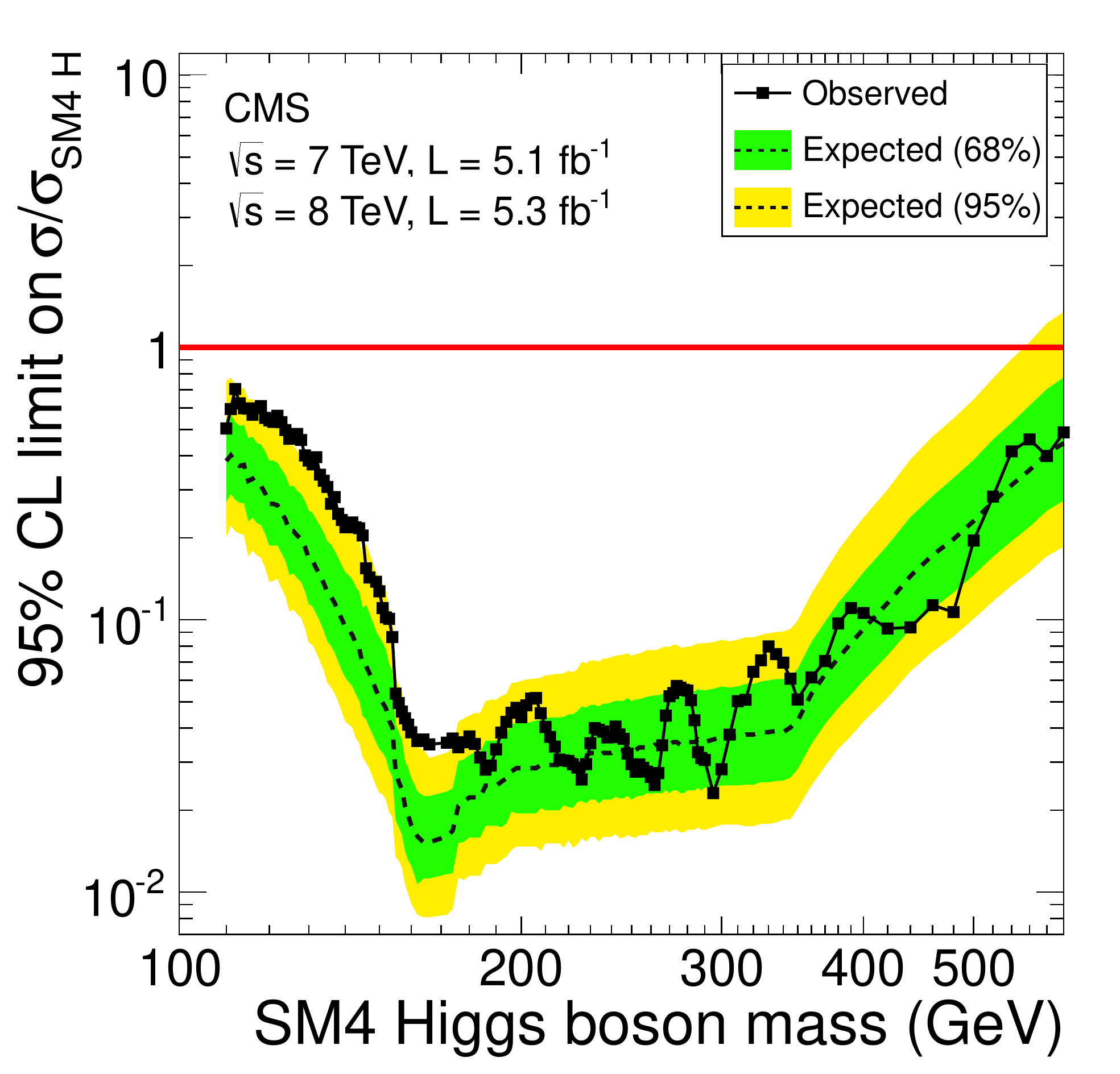}
\caption{CMS exclusion limit for the SM4 model using Higgs boson decays to $WW$, $ZZ$ and $\tau\tau$~\protect\cite{fp_sm4_cms}. 
The limit is given in units of the expected cross section in the investigated SM4 benchmark scenario.}
\label{fig-sm4}
\end{figure}

CMS searches for Higgs bosons in SM-like models with a fourth generation of fermions using up to 10~fb$^{-1}$ of 7 TeV and 8 TeV data~\cite{fp_sm4_cms}.
In the investigated benchmark scenario, the effective Higgs coupling to gluons is enhanced by a factor of 4-9 (depending on $m_H$) while 
the decay to two photons is almost entirely suppressed. The search thus focuses on Higgs bosons produced in $gg$-fusion and decaying 
to $WW$, $ZZ$ and $\tau\tau$. No significant excess compatible with the signal hypothesis is observed and when combining the 
channels considered, consequently the SM4 benchmark scenario is excluded for Higgs boson masses between 110~GeV and 600~GeV (see figure~\ref{fig-sm4}).

In fermiophobic models, the Higgs coupling to fermions is suppressed. The CMS study~\cite{fp_sm4_cms}, using 10~fb$^{-1}$ of 7 TeV and 8 TeV data, 
thus focuses on Higgs boson decays to photons in 9 different categories. The analysis excludes a fermiophobic Higgs boson with a mass in the range 
110-147~GeV.

%\clearpage

%\vspace*{0.9cm}
\section{Conclusions}
A large number of BSM Higgs boson searches have been performed by the ATLAS and CMS experiments. No significant excess is observed, 
and various cross-section limits and exclusion regions for the parameter space of several models have been provided. 
In particular the MSSM is becoming heavily constrained: A new lower limit of $m_A>125$~GeV has been set for the $m_h^\mathrm{max}$ 
scenario 
(when including the region excluded by LEP), and for $125 < m_A$ [GeV] $< 225$ only a $\tan \beta$ of
about 2-5 is still allowed. The limits extend up to $m_A=800$~GeV and $\tan\beta=50$. 
Strong cross section limits have been set for the existence of additional Higgs bosons with SM-like properties covering a 
large range of masses up to 1 TeV.
Furthermore, general 2HDMs, the NMSSM as well as type-II seesaw mechanism models have been constrained.

Very few of these analyses have analysed the full Run-I LHC dataset, and in addition predictions of several well-motivated 
BSM models have not been experimentally tested at the LHC. It is thus still possible that BSM Higgs physics is hiding in 
the current data -- and certainly, additional data collected at $\sqrt{s}=13$ TeV or higher at the LHC beginning from 2015 will 
greatly enhance the sensitivity to BSM Higgs bosons.

%
%\clearpage
%
%\textit{
%This extra page is somehow caused by the line-numbers package. Will automatically disappear once I upload a draft without line numbers.
%}
%
\end{document}